\documentclass[aps,prb,twocolumn,footinbib,showpacs,amsmath,amssymb]{revtex4-1} 

\usepackage{graphicx}
\usepackage{subfigure}
\usepackage{epsfig} 
\usepackage{dcolumn}
\usepackage{bm}
\usepackage{color}

\begin{document}

\title{Heat transfer statistics in extreme-near-field radiation}

\author{Gaomin Tang}
\email{phytg@nus.edu.sg}
\author{Jian-Sheng Wang}
\affiliation{Department of Physics, National University of Singapore, Singapore 117551, Republic of Singapore}

\date{\today}

\begin{abstract}
We investigate the full counting statistics of extreme-near-field radiative heat transfer using nonequilibrium Green's function formalism. In the extreme near field, the electron-electron interactions between two metallic bodies dominate the heat transfer process. We start from a general tight-binding electron Hamiltonian and obtain a Levitov-Lesovik like scaled cumulant generating function (SCGF) using random phase approximation to deal with electron-electron interaction. The expressions of heat current and its fluctuation (second cumulant) are obtained from the SCGF. The fluctuation symmetry relation of the SCGF is verified. In the linear response limit (small temperature gradient), we express the heat current cumulant by a linear combination of lower order cumulants. The heat current fluctuation is $2k_B T^2$ times the thermal conductance with $T$ the average temperature in the linear response limit, and this provides an evaluation of heat current fluctuation by measuring the thermal conductance in extreme-near field-radiative heat transfer.
\end{abstract}

\maketitle

\section{Introduction}
Heat transfer between two bodies in the far-field regime can be well-described by Planck's theory of black-body radiation \cite{Planck}. During the 1970s, experiments in the near field have shown that heat transfer becomes much larger than that being predicted by Stefan-Boltzmann law with gap sizes smaller than Wien's wavelength \cite{near1, near2}. Polder and van Hove (PvH) pioneered to give a theoretical description of near-field radiation \cite{PvH} using Rytov's formulation of fluctuating electrodynamics \cite{Rytov1, Rytov2, Rytov3}. In the PvH theory, the contributions of heat transfer are mainly from evanescent modes which vanish in the far field. Experimentalists have reduced the gap sizes from orders of $1\, \mathrm{\mu m}$ \cite{XJB, mu-m1, mu-m2} to several tens of nanometers, resulting in heat transfer enhancement from several folds to thousands of folds compared to the corresponding far-field results \cite{10nm-1, 10nm-2, 10nm-3, review1, review2, review3, review4}. And these experimental results can be well predicted by fluctuating electrodynamics. 
Researchers now can reduce gap sizes within a few nanometers \cite{nm1, nm2, nm3, nm4, nm5-PvH, angstrom1, angstrom2} or even down to few \textup{\AA}ngstr\"{o}ms \cite{angstrom1,angstrom2}, and study the extreme-near-field radiative heat transfer (eNFRHT). 
In this extreme near field, the propagating field represented by the vector potential is not important and heat transfer is dominated by the scalar potential, i.e., the instantaneous Coulomb interaction. There have been several works on this \cite{Yu, Mahan, JS11, JS12, JS13, JS14, JS15}, including using the formalism of nonequilibrium Green's function (NEGF) to deal with heat radiation mediated by electron-electron interaction \cite{JS11, JS12, JS13, JS14, JS15} or dipole-dipole interaction \cite{angstrom1, Xiong}. Analytical results for near-field heat radiation beyond the dipolar effects have also been presented. \cite{beyond_dipolar1, beyond_dipolar2} 

	Electronic current fluctuations in mesoscopic conductors have received intensive investigations and are very important to characterize the correlations in quantum transport \cite{Blanter}. In order to fully characterize a quantum transport process, people usually employ the formalism of full counting statistics (FCS) which yields not only average current and current fluctuation (the second cumulant), but also the higher order cumulants \cite{Levitov1, Levitov2, Levitov3, Nazarov, Tobiska, Forster, Flindt1, RMP, Flindt2, Flindt3, JS1, JS2, JS3, JS4, gm1, gm2, Ruben1, Ruben2, FuBin, Bijay, gm3, gm4, gm5}. 
FCS for heat and electronic transport in mesoscopic conductors has many applications. For example, entanglement entropy can be accessed by series of the charge cumulants \cite{entropy1, entropy2}. Gallavotti-Cohen symmetry of the generating function in FCS can reveal the symmetry of a nonequilibrium systems and can gives the fluctuation theorem of a physical quantity\cite{symmetry1, Keiji1, Keiji2, symmetry2, symmetry3}. 
Analogously, due to both thermal and quantum fluctuations, radiative heat transfer between two bodies is stochastic in nature and subject to fluctuations as well. 
The fluctuation of heat flux of black-body radiation in the far-field regime has been studied by Einstein in 1909 \cite{Einstein}, and fluctuation theorem of black-body radiation has also been recently reported \cite{FT-bb}. In the near-field regime, fluctuations of radiative heat transfer have been investigated using the fluctuating electrodynamics \cite{fluct}. A full-counting statistics investigation of the near-field heat transfer is yet to be studied, and the lacuna shall be filled. 

	In this work, we investigate the heat transfer statistics in the extreme near field dominated by the electron-electron interaction between two metallic bodies. 
Since obtaining generating function using NEGF for heat conduction have been extensively reported \cite{JS1, JS2, JS3, JS4, gm4, gm5}, we adopt the NEGF formalism which has been used to study heat current in the near-field heat radiation \cite{JS11, JS12, JS13, JS14} to study FCS. The formalism of NEGF can also gives an atomistic description of a system. 
We start from a general tight-binding Hamiltonian in the presence of Coulomb interaction and obtain the partition function using path integral in the time domain. Random phase approximation (RPA) is employed in order to deal with the Coulomb interaction. By introducing a counting parameter, we obtain the modified Hamiltonian together with the modified evolution operator. Generating function is obtained by involving the partition function with counting field and then the normalization condition. The scaled cumulant generating function (SCGF) is expressed in the energy domain and is reminiscent of Levitov-Lesovik's formula \cite{Levitov1, Levitov2, Levitov3}. From the SCGF, one can get the average heat current, the fluctuations, and even higher order cumulants. The fluctuation symmetry in the heat radiation system is verified, and one can also relate the heat current fluctuation with the thermal conductance in the linear response limit. In the numerical section, using a simple two-dot model, we show the relative difference of the current fluctuation evaluated in the linear response limit and its corresponding value at finite temperature differences and gap distances.

\section{Theoretical formalism}
\subsection{Model and Hamiltonian}

\begin{figure}
\includegraphics[width=2.8in]{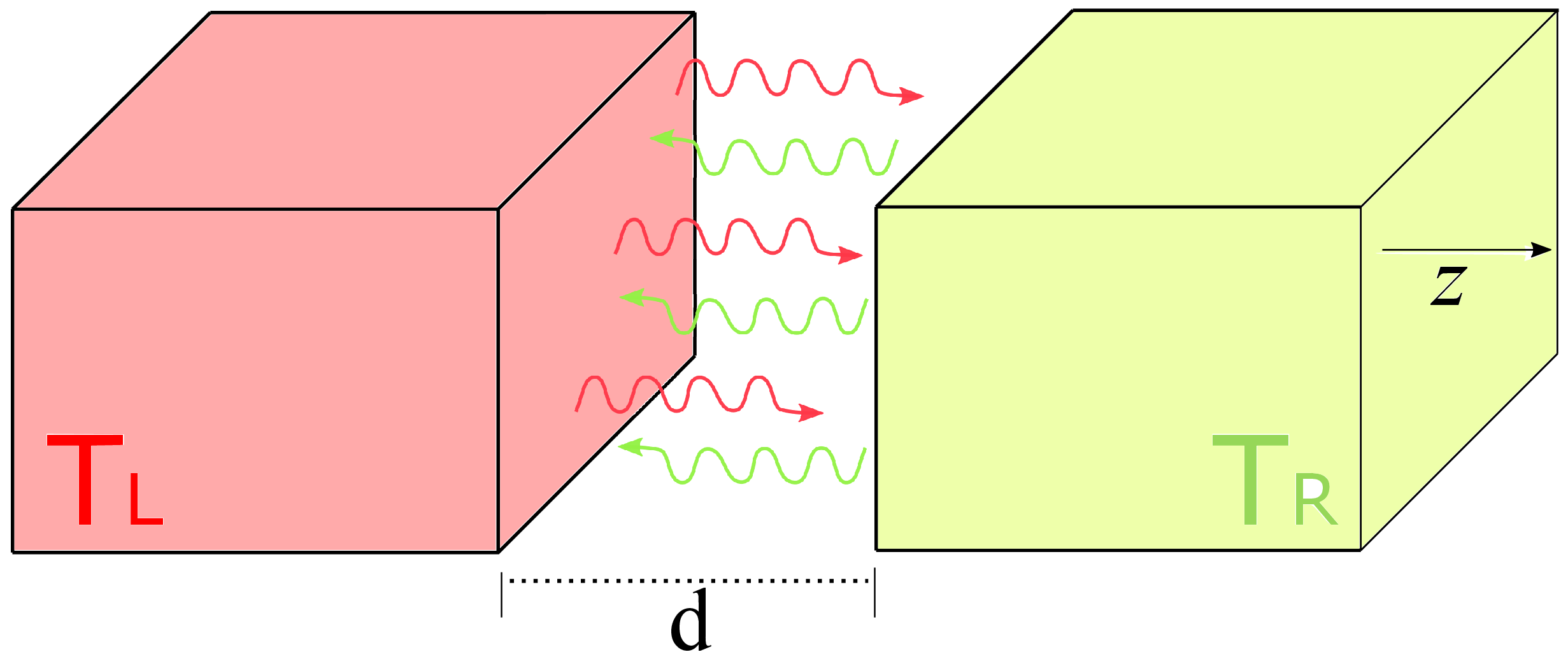} \\
\caption{Model for extreme-near-field radiative heat transfer between two vacuum-gapped semi-infinite sides meditated by Coulomb interaction. }
\label{fig1}
\end{figure}

For the extreme-near-field-radiative heat transfer system, we consider two parallel aligned cubic lattices described by tight-binding Hamiltonian (See Fig. \ref{fig1}). The two sides are maintained local thermal equilibria with different temperatures, and they exchange heat through the vacuum gap with distance $d$ via electron-electron interaction. 
The roughness of the surfaces can also be taken take care of here, since the formalism presented below is atomistic. 

	One can partition the system Hamiltonian as
\begin{equation}
H = H_{0L} + H_{0R} + V_{L} + V_{R} + V_{LR} ,
\end{equation}
where 
\begin{align}
H_{0\alpha} =& \sum_{m\in \alpha, n\in \alpha} c_m^\dag h_{mn} c_n , \label{H0} \\
V_{\alpha} =&\frac{e_0^2}{2} \sum_{m\in \alpha, n\in\alpha} c_m^\dag c_m v_{mn} c_n^\dag c_n ,\label{V} \\
V_{LR} =& e_0^2 \sum_{m\in L, n\in R} c_m^\dag c_m v_{mn} c_n^\dag c_n , \label{VLR}
\end{align}
with $\alpha = L(R)$ representing the left (right) side and $e_0$ the elementary charge. $H_{0\alpha}$ is the non-interacting Hamiltonian and $V_\alpha$ is the Coulomb interaction in side $\alpha$. $V_{LR}$ is the Coulomb interaction between the electrons on the left and right side. The front coefficient $1/2$ in $V_\alpha$ is to avoid the double counting. 
$c_m^{(\dag)}$ is the annihilation (creation) operators on the left or right side. $h_{mn}$ is the on-site energy for $m=n$ and hopping constant for $m\neq n$. 
The Hamiltonian can also be written in a compact form,
\begin{equation}
H = \sum_{mn} c_m^\dag h_{mn} c_n+ \frac{e_0^2}{2}\sum_{mn} c_m^\dag c_m v_{mn} c_n^\dag c_n .
\end{equation}
Throughout this work, the left side is set warmer than the right side so that $T_L >T_R$ with $\Delta T = T_L - T_R$.

\subsection{Partition function}
We assume that the Coulomb interaction between left and right side is absent at time $t=0$, so that the initial density matrix of the whole system at $t=0$ is the direct product of each subsystem and expressed as $\rho(0) = \rho_L \otimes \rho_R$. After time $t=0$, the interaction between left and right is turned on and the system evolves to time $t$ under the evolution operator $U(t,0) =\mathbb{T}\exp\big[ -i\int_0^t H(t')dt'/\hbar \big]$, where $\mathbb{T}$ is the time ordering operator on the Keldsyh contour. Since we let $t$ go into infinity and consider the steady state of heat transfer between two bodies, the initial system state does not influence any steady state physical quantities. 
The partition function of the whole system without any souce field or counting field is written as $Z(t)={\rm Tr} [\rho(0) U^\dag(t,0)U(t,0)] /{\rm Tr} \rho(0)$ and is exactly $1$. In the next subsection, the generating function is obtained by considering the counting field in the partition function. Using path integral on the Keldysh contour, the partition function can be expressed as \cite{Ka2}
\begin{equation}
Z(t) = \frac{1}{{\rm Tr} \rho(0)}\int {\cal D}[\bar{\phi}\phi] \exp[{i{\cal S}_{0} + i{\cal S}_{int}}] ,
\end{equation}
with ${\cal S}_{0}$ representing the action of free electron lattice·
\begin{equation}
{\cal S}_{0} = \int_C d\tau \sum_{mn}\bar{\phi}_{m} G_{mn}^{-1} \phi_{n} ,
\end{equation}
and ${\cal S}_{int}$ the Coulomb interaction
\begin{equation}
{\cal S}_{int} =-\frac{e_0^2}{2}\int_C d\tau \sum_{mn} \bar{\phi}_{m} \phi_{m} v_{mn}\bar{\phi}_{n} \phi_{n}.
\end{equation}
In the above expressions, $\bar{\phi}_{m}$ and $\phi_{m}$ are the fermionic Grassmann variables, and $G^{-1}$ is the inverse electronic Green's functions \cite{Ka2}. We have set $\hbar$ to 1. The integration is over the Keldysh contour $C$. From now on, the time on the Keldysh contour is denoted as Greek letters, and real time using Latin letters. 

	Performing the Hubbard-Stratonovich transformation \cite{Hubbard, Ka2} by introducing the real scalar field $\Psi_m$, one can reduce the four-particle interaction exactly in terms of an effective electron-photon interaction and has the expression,
\begin{widetext}
\begin{equation}
\exp(i{\cal S}_{int}) = \int {\cal D}[\Psi] \exp\bigg\{ i\int_C d\tau \bigg[ \frac{1}{2}\sum_{mn} \Psi_m \Big(\frac{v^{-1}}{e_0^2}\Big)_{mn} \Psi_n -\sum_{m}\Psi_m \bar{\phi}_{m} \phi_{m} \bigg] \bigg\} ,
\end{equation}
\end{widetext}
where $v^{-1}$ is the inverse Coulomb interaction matrix. 
The integration measure $\int {\cal D}[\Psi]$ is normalized such that $\int {\cal D}[\Psi] \exp\big\{ i\int_C d\tau \frac{1}{2} \Psi (v^{-1}/e_0^2) \Psi \big\}=1$. 
The partition function $Z(t)$ could be further simplified by integrating out the fermionic Grassmann variables with the following relation, 
\begin{align}
&\int {\cal D}[\bar{\phi}\phi] \exp \bigg[ i\int_C d\tau \sum_{mn}\bar{\phi}_{m} G_{mn}^{-1} \phi_{n} -i\sum_{m}\Psi_m \bar{\phi}_{m} \phi_{m} \bigg]  \notag \\
=& \det \big(-iG^{-1} + i\Psi\big).
\end{align}
where $\Psi = {\rm diag}[\Psi^{+}, -\Psi^{-}]$ is diagonal in Keldysh, time and lattice space. Using $\det M=\exp[{\rm Tr}\ln M]$, 
the partition function $Z$ has the form as
\begin{equation}
Z \simeq \int {\cal D}[\Psi] \exp\bigg\{ i\int_C d\tau {\rm Tr}\bigg[ \frac{1}{2}\Psi \Big(\frac{v^{-1}}{e_0^2}\Big) \Psi -i\ln(1- G\Psi) \bigg] \bigg\}  .
\end{equation}
where some front coefficients have been ignored at this moment and would be taken into consideration by using normalization condition of generating function when discussing FCS in the next subsection. 
Using the relations ${\rm Tr}\ln (1-M)= -\sum_{j=1} M^j/j$ and ${\rm Tr}(G\Psi)=0$ from the fact $G_{mn}^{++}(t_1,t_1)=G_{mn}^{--}(t_1,t_1)$ \cite{Ka2}, one can perform random phase approximation (RPA), i.e., expanding partition function to the second order of scalar field and obtain 
\begin{equation}
Z \simeq \int {\cal D}[\Psi] \exp\bigg\{ \frac{i}{2}\int_C d\tau \ {\rm Tr} \Big[ \Psi \Big(\frac{v^{-1}}{e_0^2}\Big) \Psi +i (G\Psi G\Psi) \Big] \bigg\} .
\end{equation}
Owing to the fact that $\Psi$ is diagonal, we have
\begin{align}
{\rm Tr} (G\Psi G\Psi) = {\rm Tr} \left[ \Psi
\begin{pmatrix}
G^{++}G^{++} & G^{+-}G^{-+}  \\  G^{-+}G^{+-}  & G^{--}G^{--}
\end{pmatrix}  
\Psi \right].
\end{align}
Then we can rewrite the partition function in the form as
\begin{equation}
Z \simeq\int {\cal D}[\Psi] \exp\bigg\{ \frac{i}{2}\int_C d\tau \ {\rm Tr}\bigg[\Psi \Big(\frac{v^{-1}}{e_0^2}\Big) \Psi -\Psi \Pi \Psi \bigg] \bigg\},
\end{equation}
by introducing the photon self-energy $\Pi$. Keldysh components of the photon self-energy have the expressions
\begin{equation}
\Pi_{mn}^{ab}(t_1, t_2)= -i e_0^2 G_{mn}^{ab}(t_1-t_2) G_{nm}^{ba}(t_2-t_1) ,
\end{equation}
with $a,b=+,-$ and $m,n$ belonging to the same side.
We also adopt commonly used notations $\Pi^{<,>}$ to denote the lesser and greater photon self-energy with the identities $\Pi^< \equiv \Pi^{+-}$ and $\Pi^> \equiv \Pi^{-+}$. 

	Finally, by integrating out the real scalar field, we get the partition function in the form of a Fredholm determinant in the time domain,
\begin{align}
Z(t) &\simeq \frac{1}{\sqrt{\det (v^{-1}- \Pi)}} \notag \\
&= \left\{\det \left[ 
\begin{pmatrix}
v_L & v_{LR}  \\  v_{RL}  & v_R
\end{pmatrix}^{-1}  - 
\begin{pmatrix}
\Pi_L & 0 \\  0  & \Pi_R
\end{pmatrix} \right]\right\}^{-1/2}  ,
\end{align}
where the determinant is over both the contour time and lattice space. In the above formula, $v_\alpha$ is the Coulomb interaction on the same side, and $v_{LR}=v_{RL}$ is the Coulomb interaction between the left and right side. The Dyson equation of photon Green's function is expressed as $D^{-1}=v^{-1}-\Pi$ or
\begin{equation} \label{Dyson}
\begin{pmatrix}
D_{LL} & D_{LR}  \\  D_{RL}  & D_{RR}
\end{pmatrix}^{-1}
=
\begin{pmatrix}
v_L & v_{LR}  \\  v_{RL}  & v_R
\end{pmatrix}^{-1} 
 - 
\begin{pmatrix}
\Pi_L & 0 \\   0  & \Pi_R
\end{pmatrix} .
\end{equation}
The Dyson equation in this form holds in both time and energy space.

\subsection{Full counting statistics}
The statistical behaviors of the heat transfer in a specific side are all encoded in the probability distribution $P(\Delta \epsilon, t)$ of the transferred energy $\Delta \epsilon = \epsilon_t - \epsilon_0$ between an initial time $t=0$ and time $t$.
The generating function ${\cal Z}(\lambda, t)$ with the counting field $\lambda$ is defined as,
\begin{equation} \label{Z}
{\cal Z}(\lambda,t) \equiv \langle e^{i\lambda \Delta \epsilon}\rangle = \int P(\Delta \epsilon, t) e^{i\lambda \Delta \epsilon} d\Delta \epsilon.
\end{equation}
To investigate statistical behaviors of the transferred energy from the left side, we could focus on the energy operator which is actually the free Hamiltonian $H_{0L}$.
Under the two-time measurement scheme \cite{RMP, twotime}, generating function of transferred energy can be expressed over the Keldysh contour as \cite{RMP, JS3, JS4, gm2},
\begin{align} \label{Z2}
{\cal Z}(\lambda,t) &=\mathrm{Tr}\left\{\rho(0)\mathcal{T}_C\exp\left[-\frac{i}{\hbar}\int_C H_\gamma(t')dt' \right]\right\} \Bigg/ {\rm Tr} \rho(0) \notag \\
&={\rm Tr}\left\{ \rho(0) U^\dag_{\lambda/2}(t,0) U_{-\lambda/2}(t,0) \right\} \Big/{\rm Tr} \rho(0),
\end{align}
with the modified evolution operator,
\begin{equation}\label{U}
U_\gamma(t,0) =\mathbb{T} \exp\left[ -\frac{i}{\hbar}\int_{0}^{t} H_\gamma(t') dt'\right] ,
\end{equation}
where $\gamma =-\lambda/2$ on the forward contour branch, and $\gamma = \lambda/2$ on the backward contour branch. Here the modified evolution operator is expressed by the modified Hamiltonian,
\begin{align}  \label{mH}
H_\gamma &= e^{i\gamma H_{0L}} H e^{-i\gamma H_{0L}} \notag \\
&= \widetilde{H}_{0L} + H_{0R} + \widetilde{V}_{L} + V_{R} +\widetilde{V}_{LR}.
\end{align}
The tilde over the Hamiltonians means that the annihilation (creation) operator $c_m^{(\dag)}$ on the left side in Eqs.~\eqref{H0}-\eqref{VLR} are replaced with $c_m^{(\dag)}(t_\gamma)$ with $t_\gamma=\hbar\gamma$, and $c_m(t_\gamma) = e^{i\gamma H_{0L}} c_m e^{-i\gamma H_{0L}}$ for $m\in L$. This replacement only affects the electronic Green's function on the left side with a time shift for lesser and greater component in the partition function, which means $G_{mn}^{ab}(t_1-t_2) \rightarrow G_{mn}^{ab}(t_1-t_2-(a-b)\lambda /2)$ with $m,n\in L$.

	Considering the counting field and the normalization condition, we arrive at the generating function being expressed as
\begin{align}
{\cal Z}(t) = \frac{ \sqrt{\det (v^{-1}- \Pi)} }{ \sqrt{\det (v^{-1}- \widetilde{\Pi})} } , 
\end{align}
with the transformed photon self-energy on the left side as 
\begin{align} \label{Pi_time}
&\widetilde{\Pi}_{mn}^{ab}(t_1, t_2) = \notag \\
& -i e_0^2 G_{mn}^{ab}(t_1-t_2-(a-b)\frac{\lambda}{2}) G_{nm}^{ba}(t_2-t_1-(b-a)\frac{\lambda}{2}) ,
\end{align}
for $m,n\in L$.
The photon self-energy on the right side remains unchanged. 
By using the Dyson equation, Eq.~\eqref{Dyson} and defining $\bar{\Pi}=(\widetilde{\Pi}-\Pi)$, we have 
\begin{equation} \label{GF1}
{\cal Z}(t) = \frac{1}{ \sqrt{\det \big[ 1-D_{LL}\bar{\Pi}_L \big]} }.
\end{equation}
By letting $\lambda=0$, we get a vanishing $\bar{\Pi}$ and can verify the normalization condition of the generating function.
In the long time limit, we can Fourier transform Eq.~\eqref{GF1} into energy domain with the form 
\begin{equation} \label{GF2}
\lim_{t\rightarrow \infty}\ln{\cal Z}(t) = -\frac{t}{2} \int \frac{d\omega}{2\pi} {\rm Tr}\ln\big[ 1-D_{LL} \bar{\Pi}_L \big] .
\end{equation}
The Keldysh space dimension can be eliminated by writing the term ${\rm Tr}\ln(1-M)$ as $\ln\det(1-M)$ and then using the identity (assume $A$ is invertible)
\begin{equation}
\det \begin{pmatrix}
A & B \\ C & D
\end{pmatrix} = \det A \det(D-CA^{-1}B).
\end{equation}
The scaled cumulant generating function (SCGF) ${\cal F}(\lambda)=\lim_{t\rightarrow \infty}\ln{\cal Z}(t)/t$ is expressed as 
\begin{align} \label{SCGF0}
{\cal F}(\lambda) = -\frac{1}{2} \int \frac{d\omega}{2\pi} {\rm Tr} \ln \big[ 1- & D_{LR}^r \Pi_R^< D_{RL}^a \bar{\Pi}_L^>   \notag \\
 - & D_{LR}^r \Pi_R^> D_{RL}^a \bar{\Pi}_L^< \big] ,
\end{align}	
with $\bar{\Pi}_L^{</>} = \widetilde{\Pi}_L^{</>} - \Pi_L^{</>}$. 

	The Fourier transformation of Eq.~\eqref{Pi_time} enables us to obtain the expression of transformed self-energy in energy domain as 
\begin{equation}
\widetilde{\Pi}_{mn}^{ab}(\omega)= -i e_0^2 \int \frac{dE}{2\pi} G_{mn}^{ab}(E) G_{nm}^{ba}(E-\hbar\omega) e^{i(a-b)\lambda\hbar\omega /2} ,
\end{equation}
where $E$ is in the unit of energy and $\omega$ the angular frequency. 
In the local equilibrium approximation (LEA), the electrons are maintained in an equilibrium state, so that $G_{mn}^<(E)=i A_{mn}(E) f_\alpha(E)$ and $G_{mn}^>(E)=i A_{mn}(E) [f_\alpha(E)-1]$ with the electronic spectral function $A_{mn}(E)=-2{\rm Im}(G_{mn}^r(E))$ and $m,n\in \alpha$. Using the relation 
$f_\alpha(E) [f_\alpha(E-\hbar\omega)-1] = N_\alpha(\omega) [f_\alpha(E) - f_\alpha(E-\hbar\omega)]$ with the Bose-Einstein distribution $N_\alpha(\omega) = 1/[e^{\beta_\alpha \hbar\omega}-1]$ at
the temperature $T_\alpha =1/(k_B\beta_\alpha)$, one has
\begin{equation} \label{111}
\Pi_{mn}^< (\omega) = -i N_\alpha(\omega) A_{\Pi mn} (\omega) ,
\end{equation} 
with 
\begin{align} \label{222}
A_{\Pi mn} (\omega)= e_0^2\int \frac{dE}{2\pi} &[f_{\alpha}(E)-f_{\alpha}(E-\hbar\omega)] \times \notag \\
&A_{mn}(E) A_{nm}(E-\hbar\omega) .
\end{align}
Then the lesser and greater photon self-energy can be written as 
\begin{align}
\Pi_\alpha^< (\omega) &= N_\alpha(\omega) [\Pi_\alpha^r (\omega)-\Pi_\alpha^a (\omega)] = 2i N_\alpha(\omega) {\rm Im}[\Pi_\alpha^r (\omega)] , \label{333} \\
\Pi_\alpha^> (\omega) &= [N_\alpha(\omega)+1] [\Pi_\alpha^r (\omega)-\Pi_\alpha^a (\omega)] . \label{444}
\end{align} 
From Eqs.~\eqref{111}, \eqref{222} and \eqref{333}, the retarded photon self-energy is obtained with the form 
\begin{align}
\Pi^r_{mn}(\omega) =-i e_0^2\int \frac{dE}{2\pi} \Big[ & G_{mn}^r(E) G_{nm}^<(E-\hbar\omega) \notag \\
 + & G_{mn}^<(E) G_{nm}^a(E-\hbar\omega) \Big]. \label{555}
\end{align}

	We finally arrive at the expression of SCGF as
\begin{widetext}
\begin{equation} \label{SCGF}
{\cal F}(\lambda)= -\int_0^{\infty} \frac{d\omega}{2\pi} \ln \Big\{ 1- {\cal T}(\omega) \left[ (e^{i\lambda\hbar\omega}-1) N_L(\omega)(1+N_R(\omega)) +(e^{-i\lambda\hbar\omega}-1) N_R(\omega)(1+N_L(\omega)) \right] \Big\} ,
\end{equation}
\end{widetext}
where the transmission coefficient is
\begin{equation}
{\cal T}(\omega) = 4{\rm Tr} \big\{ D_{LR}^r(\omega) {\rm Im}[\Pi_R^r (\omega)] D_{RL}^a(\omega) {\rm Im}[\Pi_L^r (\omega)] \big\} .
\end{equation}
The SCGF is reminiscent of Levitov-Lesovik's formula for electronic transport \cite{Levitov1,Levitov2, Levitov3}. The front coefficient $1/2$ in Eq.~\eqref{SCGF0} is missing in Eq.~\eqref{SCGF}, because the contributions from positive and negative angular frequency are the same for the heat current. 

	The $k$th cumulant of heat current $\langle\langle I_h^k \rangle\rangle$ could be calculated by taking the $k$th derivative of SCGF which is ${\cal F}(\lambda)$ with respect to $i\lambda$,
\begin{equation} \label{jth}
\langle\langle I_h^k \rangle\rangle = \frac{\partial^k {\cal F}(\lambda)}{\partial (i\lambda)^k} \bigg{|} _{\lambda=0}.
\end{equation}
The heat current (the first cumulant) bears a Caroli form \cite{JS11, JS12, JS13, JS14, JS15}
\begin{equation} \label{1st}
I_h = \int_0^{\infty} \frac{d\omega}{2\pi} \hbar\omega {\cal T}(\omega)\big[N_L(\omega)-N_R(\omega)\big] .
\end{equation}
The heat current fluctuation (the second cumulant) has the expression,
\begin{align} \label{2nd}
\langle\langle I_h^2\rangle\rangle = \int_0^{\infty} \frac{d\omega}{2\pi} 
(\hbar\omega)^2 \big\{ &{\cal T} [N_L(1+N_R) + N_R(1+N_L) \big]  \notag \\
+ &{\cal T}^2 \big[ N_L-N_R ]^2 \big\} .
\end{align}

	Applying the relation $N_R(1+N_L)=\exp(\Delta\beta\hbar\omega) N_L(1+N_R)$ with $\Delta\beta =\beta_L-\beta_R$ in Eq.~\eqref{SCGF}, we can verify the following fluctuation symmetry relation,
\begin{equation} \label{FT}
{\cal F}(\lambda) = {\cal F}(-\lambda -i \Delta\beta) . 
\end{equation}
This symmetry relation has already been derived for heat transfer through conductors \cite{gm4, gm5, Keiji1, Keiji2}, and it is now verified for eNFRHT where heat transfer through a gap is mediated by Coulomb interaction in the RPA level. 
This symmetry implies that the backward probability of transferred energy $-\Delta \epsilon$ from cold right side to the hot left side is exponentially suppressed with respect to the forward one with the detailed balance relation (also called Gallavotti-Cohen symmetry \cite{GC-symmetry1, GC-symmetry2})
\begin{equation} \label{pp}
\frac{P(-\Delta \epsilon)}{P(\Delta \epsilon)} = \exp\left[ (\beta_L-\beta_R)\Delta \epsilon \right] ,
\end{equation}
which also implies that
\begin{equation}
\int d\Delta \epsilon P(\Delta \epsilon) = \int d\Delta \epsilon P(-\Delta \epsilon) e^{-\Delta\beta \Delta \epsilon} \equiv \langle e^{-\Delta\beta \Delta \epsilon} \rangle =1 .
\end{equation}
The above equality, Eq.~\eqref{pp}, also holds in black-body radiation in far field \cite{FT-bb}. 

	Now we consider the universal relations for heat current cumulants under a small temperature gradient which is in analogy with the universal relation for particle current cumulants \cite{Tobiska, Forster}. In the linear response regime $\Delta\beta \rightarrow 0$, the total derivative of ${\cal F}(-\lambda-i\Delta\beta)$ with respect to $\Delta\beta$ has the following expansion
\begin{equation} \label{Taylor}
\frac{d^k {\cal F}(-\lambda-i\Delta\beta, \Delta\beta)}{d \Delta\beta^k} \bigg|_{\lambda=0}
=\sum_{j=0}^k \binom{k}{j} \frac{\partial^k {\cal F}(i\lambda, \Delta\beta)}{\partial \Delta\beta^{k-j} \partial (i\lambda)^j} \bigg|_{\lambda=0} ,
\end{equation}
where we have written the dependence of $\Delta\beta$ of SCGF explicitly out in both sides. 
Since ${\cal F}(\lambda=0, \Delta\beta) = 0$, the left hand side of Eq.~\eqref{Taylor} vanishes due to  Eq.~\eqref{FT}. The last term in the summation of Eq.~\eqref{Taylor} is the $k$th derivative of the SCGF with respect to the counting field $i\lambda$, which is actually $\langle\langle I_h^k \rangle\rangle$ in the linear response limit. Then we have the relation
\begin{equation}
\langle\langle I_h^k \rangle\rangle_l = - \sum_{j=1}^{k-1} \binom{k}{j} 
\frac{\partial^{k-j} \langle\langle I_h^j \rangle\rangle_l}{\partial \Delta\beta^{k-j}} ,
\end{equation}
in which the heat current cumulant is expressed by a linear combination of lower order heat current cumulants. Here, we've added the subscript `$l$' in $\langle\langle I_h^k \rangle\rangle$ to distinguish it from the one calculated from Eq.~\eqref{jth}. 
By specifying $k=2$, we can relate the heat current fluctuation with the heat current through $\langle\langle I_h^2 \rangle\rangle_l =-2 \partial I_h /\partial\Delta\beta$, which leads to
\begin{equation} \label{2nd_l}
\langle\langle I_h^2 \rangle\rangle_l = 2k_B T^2 G_h,
\end{equation}
where average temperature $T=(T_L+T_R)/2$, and thermal conductance $G_h \equiv \partial I_h / \partial\Delta T$ with $\Delta T=T_L-T_R$.

\section{Numerical Calculation}
Current fluctuation in an electron transport system is more difficult to be experimentally measured compared to the mean current. It is expected that the heat current fluctuation is difficult to be measured as well. Linear response relation Eq.~\eqref{2nd_l} provides us an evaluation of heat current fluctuations using heat conductance. Since the relation of Eq.~\eqref{2nd_l} is obtained in the linear response limit, actual heat current fluctuation should deviate from the one evaluated in Eq.~\eqref{2nd} beyond linear response limit. To quantify the deviation, we introduce the relative difference for the heat current fluctuation
\begin{equation}
d_r = |\langle\langle I_h^2 \rangle\rangle - \langle\langle I_h^2 \rangle\rangle_l |/\langle\langle I_h^2 \rangle\rangle .
\end{equation}

For a general problem of near-field radiation between metal objects at a distance of order nanometers, one can use recursive Green's function method to get the retarded Green's function in the absence of Coulomb interaction \cite{recursive1, recursive2}. For the situation of an infinitely large surface, periodic boundary condition can be used so that one can work in the momentum space. Having obtained the Green's functions, the photon self-energies can be obtained through convolution from Eqs. \eqref{333}-\eqref{555}. Photon Green's function can be found through matrix inversion indicated by Eq. \eqref{Dyson}. To get the electron density of states in one of the surface more accurately, the Fock self-energies are incorporated in the non-interacting retarded Green's function \cite{JS13}. 
Since Thomas-Fermi screening length in metals is usually a few lattice spacing, three to five layers are enough for convergency \cite{JS14}. 
More calculation details can be found in Refs. [\onlinecite{JS13}] and [\onlinecite{JS14}].

\begin{figure}
\includegraphics[width=\columnwidth]{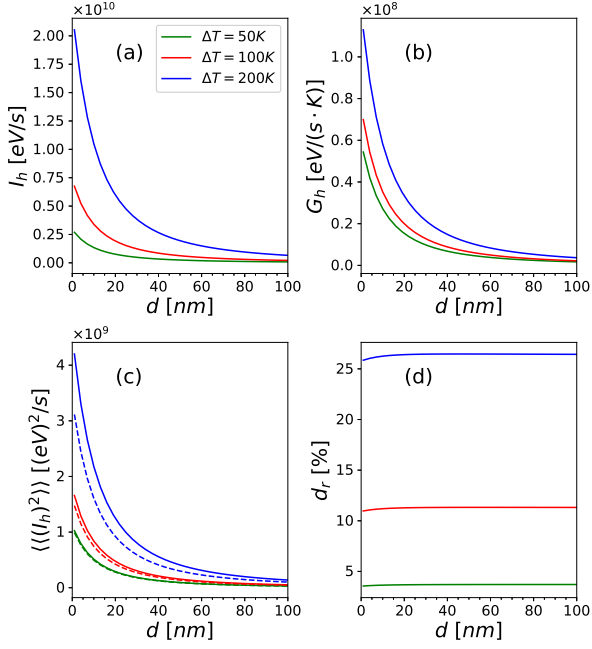} \\
\caption{(a) Heat current, (b) thermal conductance, (c) heat current fluctuation, and (d) the relative difference for heat current fluctuation versus the gap distance for different temperature gradients $\Delta T$ with $T_R=300\,$K. In panel (c), the exact heat current fluctuations and the ones in the linear response limit are plotted using solid and dashed lines, respectively. We set the chemical potentials of electron reservoirs in both sides as $\mu_L =\mu_R =0$, and quantum dot levels as $\epsilon_L=\epsilon_R=0$. Other electron reservoir constants are $\Gamma_L= 1\,$eV, $\Gamma_R= 0.5\,$eV, $E_L = 2\,$eV, and $E_R = 1\,$eV. The areas of both plates are chosen as $A=389.4\,$nm$^2$ to be close to the experimental value \cite{angstrom1}, and $Q=1.5e_0$.  }
\label{figDT}
\end{figure}

\begin{figure}
\includegraphics[width=\columnwidth]{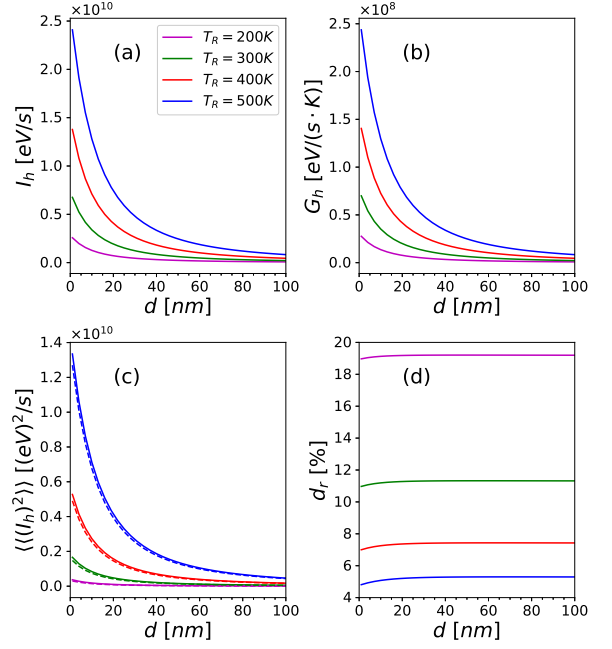} \\
\caption{(a) Heat current, (b) thermal conductance, (c) heat current fluctuation, and (d) the relative difference for heat current fluctuation versus the gap distance by varying $T_R$ with $\Delta T=100\,$K. Heat current fluctuations in the linear response limit are plotted in dashed lines in panel (c). Other parameters are the same as in Fig.~\ref{figDT}. }
\label{figTR}
\end{figure}

	For simplicity, we consider a nano-sized capacitor consisting of two quantum dots \cite{JS11}, and each plate can host a charge of $0$ or $-Q$. The retarded photon self-energy is calculated as
\begin{align}
\Pi^r_\alpha(\omega) =-i Q^2\int \frac{dE}{2\pi} \Big[ & G_\alpha^r(E) G_\alpha^<(E-\hbar\omega) \notag \\
 + & G_\alpha^<(E) G_\alpha^a(E-\hbar\omega) \Big],
\end{align}
where
\begin{align}
G_\alpha^r(E) &= [G_\alpha^a(E)]^*=  1/[E-\epsilon_\alpha -\Sigma_\alpha^r(E)] , \notag \\
G_\alpha^<(E) &= -f_\alpha(E) [G_\alpha^r(E)-G_\alpha^a(E)]  .
\end{align}
with Fermi distribution function $f_\alpha(E)=1/[\exp((E-\mu_\alpha)/(k_B T_\alpha))+1]$ at temperature $T_\alpha$ and chemical potential $\mu_\alpha$. 
The self-energies due to electron reservoirs are chosen to follow the Lorentz-Drude model \cite{Wingreen94} with $\Sigma_\alpha^r(E)=\frac{1}{2}\Gamma_\alpha/(i+E/E_\alpha)$, where $\Gamma_\alpha$ and $E_\alpha$ are electron reservoir constants.
The Coulomb interaction matrix for the capacitor is \cite{JS14}
\begin{equation}
v^{-1} = \begin{pmatrix}
C & -C \\ -C &  C
\end{pmatrix} ,
\end{equation}
where the capacitance of the parallel plate is $C=\epsilon_0 A/d$ with plate area $A$ and vacuum dielectric constant $\epsilon_0$. The photon retarded Green's function $D_{LR}^r$ is then obtained from the Dyson equation, Eq.~\eqref{Dyson}, with the form \cite{JS11},
\begin{equation}
D_{LR}^r = (D_{RL}^a)^* =[ \Pi_L^r \Pi_R^r /C - (\Pi_L^r + \Pi_R^r)]^{-1} .
\end{equation}
Heat current and fluctuation is calculated from Eq.~\eqref{1st} and Eq.~\eqref{2nd}, respectively. Thermal conductance is obtained by numerically differentiating the heat current. 

	We plot the heat current, thermal conductance, heat current fluctuation, and the relative difference for heat current fluctuation versus the vacuum gap distance by varying temperature gradient $\Delta T$ in Fig.~\ref{figDT}, and by varying $T_R$ in Fig.~\ref{figTR}. In Fig.~\ref{figDT} (c) and Fig.~\ref{figTR} (c), the exact heat current fluctuations and the ones in the linear response limit are plotted using solid and dashed lines, respectively. A non-divergent heat current is found with $d\rightarrow 0$. The $1/d^2$ divergence at short distance is the result of using a local dielectric function in the framework of fluctuating electrodynamics \cite{nm2, 2D_Dirac}. Since in our approach, we do not use such an approximation, the heat current is found to be convergent at zero distances. 
Fig.~\ref{figDT} (b) and Fig.~\ref{figTR} (b) demonstrate the behaviours of thermal conductance with respect to temperatures, that thermal conductance increases with increasing temperature gradient or average temperature. One can clearly see that the relative differences $d_r$ increase with increasing temperature gradients. The linear response approximation, Eq.~\eqref{2nd_l} becomes less accurate with decreasing average temperature, as shown in Fig.~\ref{figTR} (d). One can see from Fig.~\ref{figDT} (d) and Fig.~\ref{figTR} (d) that the relative difference is not sensitive to the vacuum gap distance.

\section{Conclusion}
In this work, using the nonequilibrium Green's function formalism, we have obtained the scaled cumulant generating function (SCGF) of the heat transfer in the extreme-near-field radiation. Random phase approximation has been used in dealing with the electron-electron interaction which meditates the heat exchange between two bodies. We have verified the fluctuation symmetry of the SCGF, and demonstrated that the probability for energy flown from the cold side to the hot one is exponentially suppressed. Both heat current and its fluctuations are obtained from the SCGF, and heat current is in a Caroli form. The heat current cumulant is shown to be expressed by a linear combination of lower order cumulants in the linear response limit. A specific case of this is that heat current fluctuation is proportional to thermal conductance in the linear response limit. We numerically show the deviations of fluctuations evaluated in the linear response limit from its value. The evaluation of fluctuation from thermal conductance becomes poorer with larger temperature gradient and lower average temperature, and the relative difference is not sensitive to the gap distance.

\begin{acknowledgements}
The authors acknowledge the financial support from RSB funded RF scheme.
\end{acknowledgements}

\end{document}